\documentclass{jfm}
\usepackage{graphicx}
\usepackage{epstopdf, epsfig}
\usepackage{siunitx}
\DeclareSIUnit{\molar}{M}

\usepackage[version=4]{mhchem}
\ExplSyntaxOn
\keys_define:nn { mhchem }
 {
  arrow-min-length .code:n =
   \cs_set:Npn \__mhchem_arrow_options_minLength:n { {#1} } 
 }
\ExplSyntaxOff
\mhchemoptions{arrow-min-length=1em}

\DeclareSIUnit{\litre}{l}
\usepackage{color}
\usepackage{comment}
\usepackage[colorlinks,citecolor = blue, linkcolor=red,hyperindex,CJKbookmarks]{hyperref}
\usepackage{amsmath}
\usepackage{amssymb}

\newcommand\Ra{\nobreak\mbox{$\mathcal R$\hskip-0.3mm$a$}}
\newcommand\Ma{\nobreak\mbox{$\mathcal M$\hskip-0.3mm$a$}}
\newcommand\St{\nobreak\mbox{$\mathcal S$\hskip-0.3mm$t$}}
\newcommand\Sh{\nobreak\mbox{$\mathcal S$\hskip-0.3mm$h$}}
\newcommand\Sc{\nobreak\mbox{$\mathcal S$\hskip-0.3mm$c$}}
\newcommand\Pe{\nobreak\mbox{$\mathcal P$\hskip-0.3mm$e$}}
\newcommand\Reyn{\nobreak\mbox{$\mathcal R$\hskip-0.3mm$e$}}

\shorttitle{Enhanced bubble growth near solidification front}
\shortauthor{J.G. Meijer, D. Rocha, A.M. Linnenbank, C. Diddens and D. Lohse}

\title{Enhanced bubble growth near an advancing solidification front}

\author{Jochem G. Meijer\aff{1}\corresp{\email{j.g.meijer@utwente.nl}},  Duarte Rocha\aff{1}, 
  Annemarie M. Linnenbank\aff{1},  Christian Diddens\aff{1},  \and Detlef Lohse\aff{1,}\aff{2}\corresp{\email{d.lohse@utwente.nl}}}

\affiliation{\aff{1}Physics of Fluids group, Max-Planck Center Twente for Complex Fluid Dynamics,  Department of Science and Technology, Mesa+ Institute and J. M. Burgers Center for Fluid Dynamics, University of Twente, P. O. Box, 217, 7500 AE Enschede, The Netherlands.
\aff{2}Max Planck Institute for Dynamics of Self-Organization, Am Fassberg 17, 37077 Göttingen, Germany.}

\begin{document}

\maketitle

\begin{abstract}
Frozen water might appear opaque since gas bubbles can get trapped in the ice during the freezing process.
They nucleate and then grow near the advancing solidification front, due to the formation of a gas supersaturation region in its vicinity.
A delicate interplay between the rate of mass transfer and the rate of freezing dictates the final shapes and sizes of the entrapped gas bubbles. 
In this work, we experimentally and numerically investigate the initial growth of such gas bubbles that nucleate and grow near the advancing ice front.  
We show that the initial growth of these bubbles is governed by diffusion and is enhanced due to a combination of the presence of the background gas concentration gradient and the motion of the approaching front. 
Additionally, we recast the problem into that of mass transfer to a moving spherical object in a homogeneous concentration field, finding good agreement between our experimental data and the existing scaling relations for that latter problem.
Lastly, we address how fluid flow around the bubble might further affect this growth and qualitatively explore this through numerical simulations.
\end{abstract}

\section{Introduction}

While slowly freezing an aqueous suspension, the dispersed objects might either be engulfed into the ice or rejected by the advancing solidification front \citep{shangguan1992analytical,rempel2001particle, dedovets2018five,tyagi2020objects,meijer2023frozen}.  
The conditions that govern this are of significant importance in material science as the arrangement of these dispersed particles profoundly shapes
the microstructure and, consequently, the functional attributes of the solidified material \citep{deville2006freezing,deville2007ice, deville2008freeze,deville2010freeze}. 
Especially during repulsion of objects by the moving front, typically occurring at low freezing velocities, the objects tend to accumulate \citep{tyagi2021multiple,tyagi2022solute}, forming a concentration profile of the objects, that can evolve in space and time.
Similarly, since gases are soluble in a large variety of liquids, \textit{e.g.},  water, silica \citep{yokokawa1986gas}, metals \citep{shapovalov2004gasar}, and sapphire \citep{bunoiu2010gas,ghezal2012observation}, but not in their solid states, during the solidification process,  accumulation of gases at the front occurs.

For example,  when freezing water, gas bubbles nucleate at the advancing solidification front \citep{carte1961air,maeno1967air,bari1974nucleation,lipp1987investigation}, as the dissolved gases are rejected by the growing ice crystal, accumulate at the front, leading to a favourable environment for bubbles to grow.
While immiscible, 'soft' particles, such as drops,  are subjected to stresses  \citep{gerber2022stress} during the encapsulation into the ice, leading to potential deformation \citep{tyagi2021multiple,tyagi2022solute,meijer2023thin,meijer2023freezing}, the complexity during bubble entrapment is amplified due to additional mass transfer.  
The shapes and sizes of the entrapped gas bubbles in ice, ranging from very small and barely deformed to elongated vertical cylinders, are set by a delicate interplay between the rate of freezing and the rate of mass transfer, and have been studied extensively \citep{carte1961air,maeno1967air,bari1974nucleation,alley1999conditions, wei2000shape,wei2004growths,yoshimura2008growth, chu2019bubble,shao2023growth,thievenaz2024universal}.  
The initial growth of the bubbles near an advancing solidification front has however received only little attention \citep{bari1974nucleation,lipp1987investigation} and forms the main focus of this work. 
Given the presence of gradients in both gas concentration (due to accumulated gases at the front) \textit{and} temperature (governing the overall freezing process), in combination with the fact that the bubble is approached by a solid interface, a non-trivial growth of these bubbles is ensured.
In addition, complex fluid flow structures might emerge around the growing bubble, further affecting its growth.
In this work, we aim to delineate the importance of the different processes involved through a combination of experiments and numerical simulations.
We will show that this seemingly simple configuration of a bubble near an ice front underlies truly rich physics, making it a prime example of a physicochemical hydrodynamical system out of equilibrium \citep{lohse2020physicochemical}.

After introduction of the experimental and numerical procedures in section \textsection\,\ref{sec:exp}, we continue by addressing the front propagation in section \textsection\,\ref{sec:frontprop}.
We then turn to the main observation of this work which is the bubble growth near the advancing solidification front (\textsection\,\ref{sec:bubblegrowth}).
More specifically, we address the experimentally observed growth at early stages and compare the results to our numerical simulations.  
This process occurs simultaneously with the overall freezing process but at much smaller time-scales. 
In \textsection \,\ref{sec:mass transfer} we then address the enhanced mass transfer in more detail.
We end with conclusions and an outlook in section \textsection\,\ref{sec:conclusion}.
In the appendix we also discuss the potential effect of fluid flow around the growing bubble and find it to be small.

\section{Experimental procedure \&  numerical methods}
\label{sec:exp}

\subsection{Experimental procedure}

\begin{figure}
  \centering
  \includegraphics[width=0.95\textwidth]{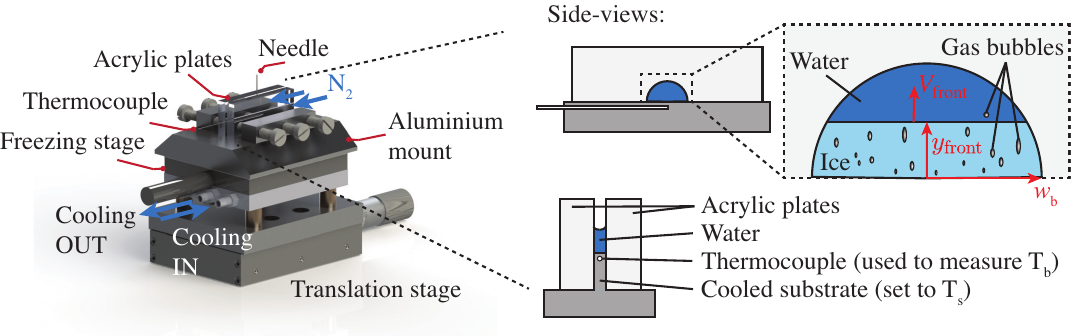}
  \caption{The experimental Hele-Shaw set-up (left) and schematic side-views of a freezing sessile drop sandwiched between the two acrylic plates (right). A thermocouple is used to measure $T_b(t)$ and the substrate temperature is fixed at $T_s$. For the analysis we distinguish between front propagation driven by thermal effects (\S\,\ref{sec:frontprop}), and the more rapid (diffusive) bubble growth near the front (\S\,\ref{sec:bubblegrowth}).}
\label{fig:1}
\end{figure}

The aim of the experimental set-up is to freeze a sessile water drop on a cold substrate.  
During the freezing process, bubbles will naturally nucleate and grow at and near the advancing solidification front.
To avoid lensing effects we are interested in freezing only a thin slice of purified water (Milli-Q), which we will refer to as 'drop' in the remainder of the text. 
In order to achieve such a quasi-two-dimensional drop, an aluminium mount is placed on top of a freezing stage (BFS-40 MPA, Physitemp) that allows two acrylic plates to be pressed against a thin metal strip (see figure\,\ref{fig:1}).
The gap between the plates is $\SI{1}{mm}$ and the temperature of the substrate close to the base of the drop, $T_b$, is measured by a thermocouple that is placed inside a groove at the side of the metal strip.
We make sure that the desired substrate temperature has been reached well within $\pm \, \SI{0.1}{\kelvin}$ for several minutes before starting the experiment (see App.\,\ref{appA}).
A needle (Nordson) and a syringe pump (PHD 2000 Infusion, Havard Apparatus) are used to deposit drops of equal volume ($V_d = \SI{25}{\micro \liter}$) between the plates, resulting in drop heights of roughly $\SI{3}{\mm}$. 
To guarantee freezing as soon as the (room temperature) water touches the substrate and to avoid supercooling, we only deposit the drop once ice crystals have formed on top of the thin metal strip.  A gentle flow of nitrogen along the outsides of the plates prevents fog and frost formation that otherwise would obscure the view.
The drop is illuminated with a diffused cold-LED to avoid local heating.
The freezing process is recorded in side-view using a camera (Nikon D850) connected to a long working distance lens (Thorlabs, MVL12X12Z). 
Once the freezing process is complete, the plates are dismounted, the frozen drop is removed and the experimental set-up is cleaned. 
Experiments are then repeated for different substrate temperatures.

Whereas the overall freezing process, driven by thermal effects, and the front propagation are extensively discussed in \S\,\ref{sec:frontprop}, the growth of gas bubbles near the front, occurring at much smaller time-scales,  forms the focus of \S\,\ref{sec:bubblegrowth}.
Before reporting on all the experimental observations, we briefly go over the technical details of our numerical simulations, for the interested reader, that are used in \textsection\,\ref{sec:bubblegrowthatfront}\,\&\,App.\,\ref{appC}, and that mimick the bubble growth near the moving front. 

\subsection{Numerical simulation}
\label{sec:numsim}
For a more detailed analysis of the growth of the bubbles near an advancing solidification front (discussed in \textsection\,\ref{sec:bubblegrowthatfront}\,\&\,App.\,\ref{appC}), we rely on axisymmetric numerical simulations that account for all the relevant physical mechanisms and allow for a comparison to the experimental observations. 
To this end, we use a sharp-interface arbitrary Lagrangian-Eulerian finite element method from the \textsc{pyoomph} package \citep{diddens2023bifurcation}, based on \textsc{oomph-lib} \citep{heil2006oomph} and \textsc{GiNaC} \citep{bauer2002introduction}. 
The domains here are represented by a triangular mesh. 

In our simulations, we consider a rectangular domain with a hole representing the bubble. 
The bubble has an initial radius of $R_0 = \SI{10}{\micro\meter}$ and is centred at the axis of symmetry, located $\SI{125}{\micro\meter}$ from the bottom boundary. 
The side walls of the domain are far enough from the bubble (25$R_0$) to neglect any boundary effects. 
We assume the dissolved gases in the water to be a single species (see \textsection\,\ref{sec:bubblegrowthatfront}). 
The gas concentration, \textit{i.e.}, the mass of dissolved gas per volume, denoted by $C$, evolves according to an advection-diffusion equation, see \eqref{Eq:AdvectionDiffusionGasC}. 
If thermal effects are considered, the temperature field $T$ is governed by the heat equation, see \eqref{Eq:HeatEquation}. 
We neglect thermophoresis effects \citep{piazza2008thermophoresis} as our tests indicate that their inclusion is of no significance.
The fluid motion in the liquid surrounding the bubble is described by the incompressible Navier-Stokes equations (see \eqref{Eq:NavierStokes}), leading to a system of equations that reads
\begin{equation}
  \partial_t C + \mathbf{u} \bcdot \bnabla C = D \bnabla^2 C,
  \label{Eq:AdvectionDiffusionGasC}
\end{equation}
\begin{equation}
  \partial_t T + \mathbf{u} \bcdot \bnabla T = \kappa \bnabla^2 T,
  \label{Eq:HeatEquation}
\end{equation}
\begin{equation}
  \bnabla \bcdot \mathbf{u} = 0 \quad \text{and } \quad \rho_\text{l} \left( \partial_t \mathbf{u} + \mathbf{u} \bcdot \bnabla \mathbf{u} \right) = - \bnabla p_l + \mu \bnabla^2 \mathbf{u},
  \label{Eq:NavierStokes}
\end{equation}
where  $\mathbf{u}$ is the fluid flow velocity, $D$ and $\kappa$ the gas and thermal diffusivity, respectively, $\rho_l$ is the liquid density, $\mu$ its dynamic viscosity, and $p_l$ the pressure in the liquid.

At the interface of the bubble, we impose a constant gas concentration, specifically the saturation concentration $C_{\mathrm{sat}}$. 
Additionally, we take the fluid velocity at the interface to be continuous and consider a dynamic boundary condition, including Marangoni stresses:
\begin{equation}
   p +  \mu \mathbf{n} \bcdot ( \bnabla \mathbf{u} + (\bnabla \mathbf{u})^T) \bcdot \mathbf{n}= \sigma \mathcal{K},
    \label{Eq:Laplace}
\end{equation}
\begin{equation}
  \mu \mathbf{n} \bcdot ( \bnabla \mathbf{u} + (\bnabla \mathbf{u})^T)\bcdot \mathbf{t} = \bnabla_{\Gamma} \sigma \bcdot \mathbf{t}.
  \label{Eq:Marangoni}
\end{equation}
Here, $p = -p_l + p_g $, with $p_l$ the pressure in the liquid and $p_g$ the gas pressure in the bubble; $\sigma(T)$ is the (temperature-dependent) surface tension, $\mathcal{K}$ the curvature of the interface, $\mathbf{n}$ and $\mathbf{t}$ the outwards-pointing normal and tangent to the bubble-liquid interface, respectively, and $\bnabla_{\Gamma}$ is the surface gradient operator.  

Despite not modelling the flow inside the bubble, it is crucial to consider the pressure exerted by the bubble on the surrounding fluid.
The mass of gas within the bubble, denoted by $m_g=\rho_g V_g$, where $\rho_g$ is the gas density and $V_\text{g}$ the bubble volume, increases due to the gas mass transfer from the surrounding supersaturated liquid, \textit{i.e.}, locally $C > C_{\mathrm{sat}}$, into the bubble through diffusion.
More specifically,
\begin{equation}
\partial_t m_g = \int_{\partial \Omega} -j{\rm d}A, \quad \text{with} \quad j=\rho_g D\bnabla C\bcdot \mathbf{n},
 \label{Eq:masstransfer}
\end{equation}
where $\partial \Omega$ is the bubble-liquid interface domain, $j$ the mass transfer rate, and $A$ the interfacial area of the bubble.
The kinematic boundary condition satisfies
\begin{equation}
 \rho_g(\mathbf{u} - \partial_t \mathbf{R})\bcdot \mathbf{n}=-j,
\end{equation} 
with $\partial_t \mathbf{R}$ the Lagrangian interface velocity.
It is important to note that $\rho_g$ may not remain constant during the bubble growth. 
We therefore assume it to evolve according to the ideal gas law,  $p_g V_g = m_g \bar{R} T_g$, where $\bar{R}$ is the gas constant, and $T_g$ the temperature within the bubble, considered constant. 

Lastly, to account for the planar solidification front that approaches the bubble during its growth, we let the bottom boundary advance with a prescribed velocity $V_{\mathrm{front}}$ (see \textsection\,\ref{sec:freezingrate}).
Its temperature is kept constant and is set to the melting temperature of water $T_m = \SI{0}{\celsius}$. 
We assume no significant action of volume-change convection at the wall \citep{davis2001theory}, \textit{i.e.}, $\mathbf{n} \cdot \mathbf{u}= V_{\mathrm{front}} \left(1 - \rho_i/\rho_l \right) = 0$, thereby disregarding any density difference between water ($\rho_l$) and ice ($\rho_i$). 
The top boundary, which is sufficiently distanced from the bubble, moves with the bottom boundary to mimic an infinite domain at each time step. 
Its temperature and gas concentration are kept constant at $T_{\mathrm{top}}$ and $C_0$, respectively (see App.\,\ref{appA}\,\&\,\ref{appB}).

The evolution of the gas pressure, volume, and mass is implemented using three Lagrange multipliers, each representing one of these quantities. 
These are coupled as described in the system of equations and solved monolithically.
While the liquid-gas interface is allowed to grow, its center position is fixed via an additional Lagrange multiplier. 
We do so, in order to account for the experimentally observed pinning of the bubble (see \textsection\,\ref{sec:bubblegrowthatfront}), which likely occurs on the acrylic plate.
This imposes a force in the liquid bulk to offset the hydrostatic pressure, ensuring the bubble remains neutrally buoyant.
Before solving the system of equations transiently, we first satisfy the stationary diffusive conditions for the gas concentration and temperature fields around the bubble. 
This is achieved by solving equations \eqref{Eq:AdvectionDiffusionGasC} and \eqref{Eq:HeatEquation} with the appropriate mentioned boundary conditions while disregarding the velocity and time-dependent terms, \textit{i.e.}, $\bnabla^2 C = 0$ and $\bnabla^2 T = 0$, respectively. 
We track the resulting bubble radius at each time step by calculating it from the value of the Lagrange multiplier representing its volume and compare with the experimental results in the subsequent sections.

\section{Front propagation}
\label{sec:frontprop}

\subsection{General experimental observations}
\label{sec:generalobs}

\begin{figure}
  \centering
  \includegraphics[width=\textwidth]{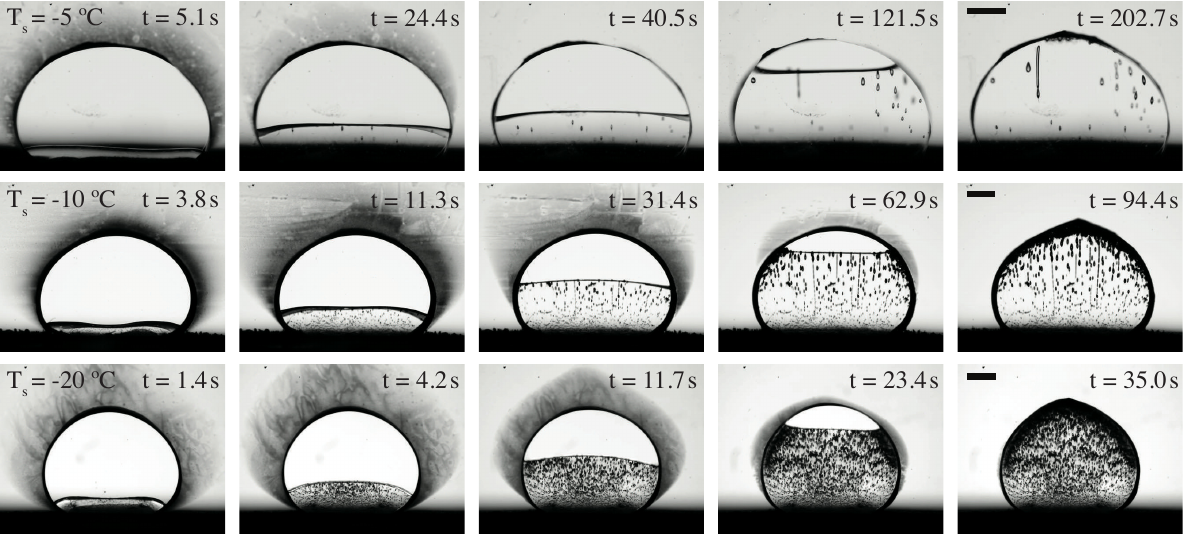}
  \caption{Sequences of images (Suppl. Movies 1-3) highlighting the differences when freezing a sessile drop of equal volume ($V_d = \SI{25}{\micro \litre}$) sandwiched between two acrylic plates at three different substrate temperatures, \textit{i.e.}, $T_s = \SI{-5}{\celsius}$, $T_s = \SI{-10}{\celsius}$ and $T_s = \SI{-20}{\celsius}$ from top to bottom, respectively. All scale bars (right panels) are $\SI{1}{\mm}$. }
\label{fig:2}
\end{figure}

When the room temperature drop is deposited onto the cooled, frosted substrate, it immediately starts to freeze from the bottom up.  
The freezing processes for three different substrate temperatures are shown as sequences of images in figure\,\ref{fig:2}. 
In all cases, we observe the immediate emergence of a shaded region around the drop that slowly fades away as the drop solidifies, which is known as a 'frost halo' \citep{jung2012frost}.
Due to the rapid phase change at very early times (shock freezing), an excess amount of released latent heat causes the drop to evaporate.
The water vapour consequently condensates at the inside of the cold acrylic plates before slowly evaporating once more.  
Apart from the significant differences in time it takes the drops to solidify ($\SI{203}{\second}, \SI{95}{\second}, \SI{35}{\second}$ for $T_s = \SI{-5}{\celsius}, \SI{-10}{\celsius}, \SI{-20}{\celsius}$, respectively), discussed in more detail in \textsection\,\ref{sec:freezingrate},  we observe that the base width of the deposited drop, $w_{\mathrm{b}}$ (see figure\,\ref{fig:1}), becomes smaller when the substrate temperature is reduced. 
As the drop is deposited from a certain height, which might slightly vary from case to case, the spreading dynamics is governed by an interplay between contact line motion and phase transition phenomena \citep{koldeweij2021initial,grivet2022contact}, leading to an earlier arrest of the contact line on colder substrates.
During the solidification process the roughly planer ice front advances at a velocity $V_{\mathrm{front}} = \mathrm{d}y_{\mathrm{front}}/\mathrm{d}t$ (see figure\,\ref{fig:1} and \S\,\ref{sec:freezingrate}).
As the water solidifies into a crystalline structure, \textit{i.e.}, ice, the gases dissolved in water are expelled and accumulate at the moving front \citep{tiller1953redistribution}.  
This lowers the threshold for bubbles to nucleate at the front, after which they start to grow. 
The initial growth of such bubbles is the main focus of the second part of this paper, see \textsection\,\ref{sec:bubblegrowth}.
The number of nucleated bubbles and their final shapes and sizes when engulfed in the ice are governed by the rate of freezing, and hence the substrate temperature (see figure\,\ref{fig:2}), as well as the locally available gas content \citep{carte1961air,maeno1967air,bari1974nucleation}.
Whereas many small bubbles are trapped in the ice when $T_s = \SI{-20}{\celsius}$, fewer and more elongated ones are formed when increasing the substrate temperature to $T_s = \SI{-10}{\celsius}$ and eventually $T_s = \SI{-5}{\celsius}$ (see right panel in figure\,\ref{fig:2}). 
At even lower freezing rates bubbles do not nucleate any more and clear ice is formed \citep{bari1974nucleation}.
Finally, at the latest stage of the freezing process, we recover the iconic pointy tip of a frozen water drop \citep{marin2014universality}, albeit less pronounced compared to the three-dimensional case due to confinement. 

\subsection{Rate of freezing}
\label{sec:freezingrate}

\begin{figure}
  \centering
  \includegraphics[width=0.85\textwidth]{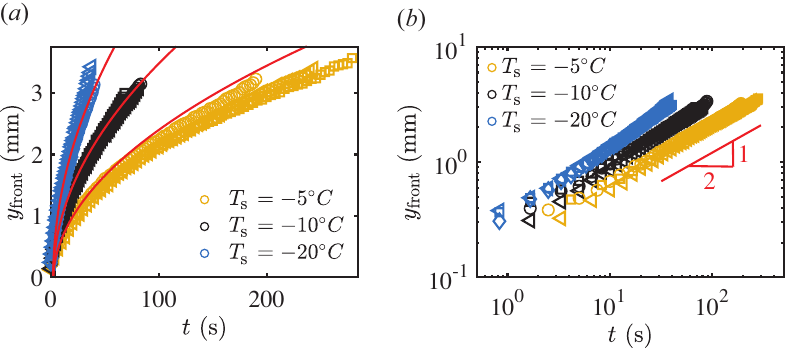}
  \caption{Position of the ice front, $y_{\mathrm{front}}$, as a function of time on linear ($a$) and double-logarithmic ($b$) scale, for repeated experiments (different symbols) at three different substrate temperatures (different colors).  The solid red lines in ($a$) correspond to (\ref{Eq:FrontEquationsTheo}).}
\label{fig:3}
\end{figure}

Before addressing the bubble growth near the ice front, we first briefly turn to the solidification dynamics of the drop, quantified by the position of the ice front $y_{\mathrm{front}}(t)$.  
For several drops at three different substrate temperatures we measure the ice thickness as a function of time and observe repeatably that, as expected, faster freezing occurs at lower substrate temperatures (see figure\,\ref{fig:3}). 
Given the quasi-two-dimensional nature of the experiments, we model the rate of freezing by simply considering the heat balance at a planar solid-liquid interface in the absence of bulk flow, which is given by Fourier's law of heat conduction \citep{davis2001theory}:
\begin{equation}
\rho_i \mathcal{L}V_{\mathrm{front}} = \left(k_i \nabla T_i - k_{\ell} \nabla T_{\ell} \right) \cdot \mathbf{n}.
\label{Eq:StefanCondition}
\end{equation}
Here, $\rho_i$ is the mass density of ice, $\mathcal{L}$ the latent heat of solidification, $k$ the thermal conductivity of ice ($i$) and water ($\ell$), respectively,  $T$ the temperature in both phases and $\mathbf{n}$ the normal vector to the interface. 
Assuming that all the heat conducted through the ice contributes to its growth and that heat conduction through the liquid can be neglected, (\ref{Eq:StefanCondition}) simplifies to
\begin{equation}
\rho_i \mathcal{L}\frac{\mathrm{d}y_{\mathrm{front}}}{\mathrm{d}t} =k_i \frac{T_m - T_b}{y_{\mathrm{front}}} = \lambda \left( T_b - T_s \right),
\label{Eq:FrontEquationsFull}
\end{equation}
where $T_m$ is the melting temperature of water, $T_b$ the bottom temperature, and $T_s$ the imposed temperature on the substrate. 
We model the heat flux from the substrate to the boundary using a heat-transfer coefficient $\lambda = k_s/d$, where $k_s$ is the thermal conductivity of the (aluminium) substrate and $d$ its (effective) thickness. 
The second equation in (\ref{Eq:FrontEquationsFull}) is solved to determine $T_b(y_{\mathrm{front}})$ (see App.\,\ref{appA}), and to extract values for $\lambda$. 
The first equation in (\ref{Eq:FrontEquationsFull}) then yields
\begin{equation}
\frac{\mathrm{d}y_{\mathrm{front}}}{\mathrm{d}t} = \frac{k_i}{\rho_i \mathcal{L}} \frac{1}{y_{\mathrm{front}}} \left[T_m - \frac{k_i T_m + \lambda T_s y_{\mathrm{front}}}{k_i + \lambda y_{\mathrm{front}}} \right],
\label{Eq:FrontEquationsTheo}
\end{equation}
which is solved numerically and compared to the experimental observations (see red lines in figure\,\ref{fig:3}\,($a$)), resulting in a good agreement when $T_s = \SI{-20}{\celsius}, \SI{-10}{\celsius}$ and an over-prediction when $T_s = \SI{-5}{\celsius}$. 
We argue that this discrepancy stems from the additional heat loss, potentially through the acrylic plates, that becomes more important at later times during slower freezing, and is not taken into account in the model.
As $y_{\mathrm{front}} \rightarrow \infty$, the right-hand term in the brackets of (\ref{Eq:FrontEquationsTheo}) approaches $T_s$ and it follows that $y_{\mathrm{front}}(t) \approx \left(2\kappa_i \St\, t \right)^{1/2}$, where $\kappa_i = k_i/(\rho_i c_p) \approx \SI{1e-6}{\meter \squared \per \second}$ is the thermal diffusivity in the ice, and where the Stefan number, $\St = \mathcal{L}/c_p(T_m - T_s)$ is defined as the ratio of latent to sensible heat, with $c_p$ the heat capacity of ice.  
At the later stages of the freezing process we converge to the appropriate scaling of $y_{\mathrm{front}} \sim t^{1/2}$ (see figure\,\ref{fig:3}($b$)).
The observed deviations between experiment and theory at the end, especially for the case where $T_s = \SI{-20}{\celsius}$, are most likely a consequence of the non-planer geometry of the drop. 

\section{Bubble growth}
\label{sec:bubblegrowth}

In this section we elaborate on the gas accumulation at the moving ice front and the initial growth of bubbles near the advancing front, which forms the main focus of this paper.

\subsection{Gas accumulation at the moving ice front}
\label{sec:gasaccumulation}

Gases are naturally dissolved in water.
As water turns into ice, these gases are expelled and accumulate at the advancing solidification front.  
Under steady conditions, at times long compared to the diffusive time $D/\left(KV_{\mathrm{front}}^2 \right)$, where $K$ is the partition coefficient,  and for constant front propagation $V_{\mathrm{front}}$, the expression of the steady concentration profile of the accumulated gases at the front reads \citep{tiller1953redistribution,pohl1954solute,carte1961air}
\begin{equation}
C(y) = C_0 \left( 1 + \frac{1-K}{K} \exp\left[-\frac{V_{\mathrm{front}}}{D} y \right] \right),
\label{Eq:ConcentrationProfile}
\end{equation}
where $D \approx \SI{1e-9}{\meter \squared \per \second}$ is the gas diffusion coefficient \citep{bari1974nucleation}, $K = 0.037 \pm 0.007$ is the partition coefficient for the gases in a water-ice system and $C_0 = \left(9 \pm 3 \right) \, \SI{}{\mg \per \litre}$ the gas concentration in water far away from the front, both determined experimentally (see App.~\ref{appB}).

Nucleated bubbles grow in this enriched environment near the moving ice front at a sufficiently high rate, as will show, that allows for the quasi-steady approximation on the front propagation in (\ref{Eq:ConcentrationProfile}). 
Additionally, any thermal effects on the bubble growth itself are neglected, as the thermal gradient in the liquid ahead of the ice front is sufficiently small (see App.~\ref{appA}). 
Given the typical size of the bubble, the water is therefore assumed to be isothermal at a value close to $T_m$, during its growth. 
Thermal effects only govern the overall freezing process of the droplet, which occur at a much larger time-scale.

\subsection{Bubble growth near the approaching ice front}
\label{sec:bubblegrowthatfront}

Bubbles nucleate at the moving ice front since the gases dissolved in water are expelled by the growing ice crystal, accumulate at the front and consequently locally lower the nucleation threshold.  
It is important to note that the content of these bubbles can differ significantly compared to the composition of air in the atmosphere ($\approx 0.79/0.21/\SI{4e-4}{}$ vol. for N${_2}$/O${_2}$/CO${_2}$).
The reason for this is the substantial difference in solubility of the individual gases in water in combination with the different partial pressures of the respective components under standard conditions \citep{lohse1997inert}. 
Using Henry's law we estimate a bubble composition that is enriched in O$_2$ and CO$_2$ compared to air ($\approx 0.59/0.38/0.03$ vol. for N${_2}$/O${_2}$/CO${_2}$), in line with experimental observations \citep{bruns1937nabliudeniia,matsuo1966gas,tsurikov1979formation}.  
For the remainder of the text we will refer to the content of the bubble as 'gas' and assume parameter values for the individual (dominant) components N$_2$ and O$_2$ (see App.\,\ref{appC}).    

\begin{figure}
  \centering
  \includegraphics[width=\textwidth]{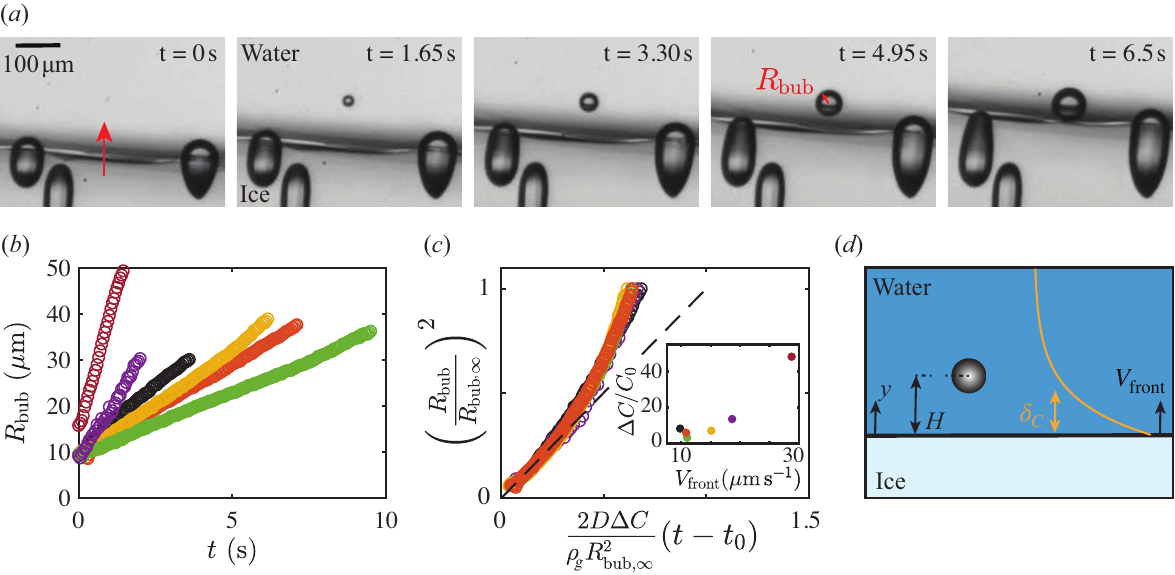}
  \caption{($a$) Sequence of images (Suppl. Movie 4) showing the growth of a bubble ahead of the solidification front, as well as ($b$) the bubble radius $R_{\mathrm{bub}}(t)$ as function of time $t$, for several different bubbles (different colors). ($c$) Bubble growth normalised by the final bubble size $R_{\mathrm{bub},\infty}$, before it is engulfed into the ice, as compared to (\ref{Eq:EpsteinPlesset}) (dashed black line).  We define $t-t_0<0$ as the time when $R_{\mathrm{bub}}$ would have been zero. The inset shows the initial concentration difference, $\Delta C$,  governing the initial growth rate (see (\ref{Eq:EpsteinPlesset})) as a function of the advancing velocity $V_{\mathrm{front}}$. ($d$) Schematic of the bubble near the ice moving front in a supersaturated environment. For all cases we find that $H \approx H_0 - V_{\mathrm{front}}t$, with $H_0$ the initial bubble-front distance, meaning that bubble centers remain stationary during their growth. The typical length of the diffusion boundary layer at the front is denoted by $\delta_C$.}
\label{fig:4}
\end{figure}

Although the majority of bubbles nucleate at the solidification front, where the rough ice surface functions as favourable nucleation site, occasionally, but not too sparsely, we observe bubble nucleation near, but not at, the advancing solidification front (see figure\,\ref{fig:4}($a$)).  
At a certain initial distance from the front a nucleation site might be present in the form of a scratch in the acrylic plate, a dust particle, or some other type of impurity, hence assuming some kind of nucleation crevice \citep{atchley1989crevice}.  The size of this initial crevice is below our optical resolution (see Suppl. Movie 4).
As the ice front approaches, and with that the accumulated gases, this nucleation site experiences a rapidly changing and gas-enriched environment,  causing it to grow into a spherical bubble, whose size is characterised by $R_{\mathrm{bub}}$ (see figure\,\ref{fig:4}($b$)). 
Initially, when $R_{\mathrm{bub}} \ll \delta_C$, where $\delta_C = D/V_{\mathrm{front}} \approx \SI{100}{\micro \meter}$ is the typical length of the diffusion boundary layer at the front (see (\ref{Eq:ConcentrationProfile})), it might be assumed that the bubble experiences a homogeneous background concentration field \citep{bari1974nucleation,lipp1987investigation}. 
During this stage, the bubble growth is dominated by diffusion and is well described by the steady Epstein-Plesset model \citep{epstein1950stability}
\begin{equation}
\frac{\mathrm{d}R_{\mathrm{bub}}}{\mathrm{d}t} = \frac{D \Delta C}{\rho_g}\frac{1}{R_{\mathrm{bub}}}, \quad \text{and thus } \quad R_{\mathrm{bub}}^2 = \frac{2D \Delta C}{\rho_g} \left(t -t_0 \right),
\label{Eq:EpsteinPlesset}
\end{equation}
with $\rho_g$ the mass density of the gas and $\Delta C = C(H_0) - C_{\mathrm{sat}}$ the initial concentration difference between that at the bubble interface and the far field. 
We define $t-t_0<0$ as the time when $R_{\mathrm{bub}}$ would have been zero and find, rather surprisingly, that for all cases the bubble-front distance $H$ (see figure\,\ref{fig:4}($d$)), linearly decreases according to the front velocity $V_{\mathrm{front}}$.
Consequently,  the centres of the bubbles considered in this work remain stationary during their growth and show no migration towards or away from the front, and do not rise due to buoyancy.
The reason presumably is some sort of pinning to the side walls. 

\begin{figure}
  \centering
  \includegraphics[width=0.95\textwidth]{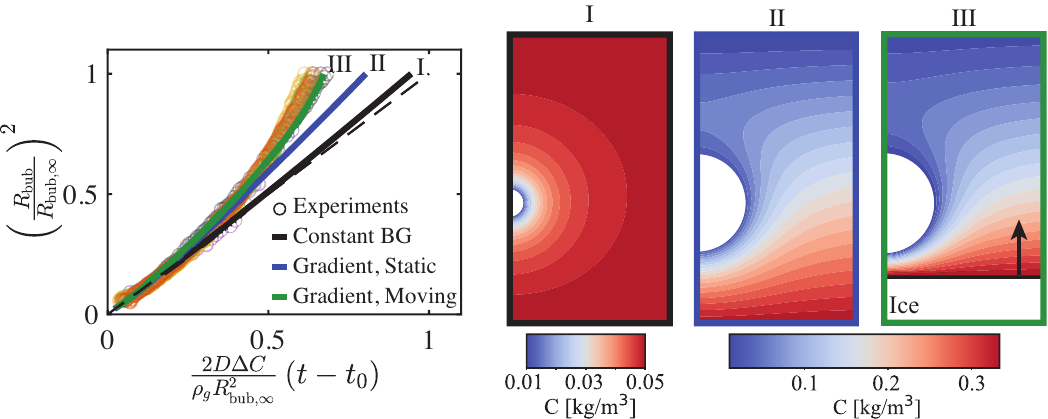}
  \caption{Comparison between experiments (data points) and numerical simulations (solid lines) (Suppl. Movie 5) on the growth of a bubble near a solidification front, namely when considering no advection for (I) a homogeneous background concentration with a static front, and for a background concentration gradient according to (\ref{Eq:ConcentrationProfile}) with a static (II) or moving (III) front.  The panels show the concentration profiles around the bubbles 5 seconds after initialisation.  The initial bubble radius is $R_0 = \SI{10}{\micro \meter}$, $C_{\mathrm{sat}} = \SI{0.029}{\kg \per \meter \cubed}$, and $V_{\mathrm{front}} = \SI{10}{\micro \meter \per \second}$ (for III only).}
\label{fig:6}
\end{figure}

The initial concentration difference, $\Delta C$, depends on (among others) the initial size of the bubble, setting the saturation concentration $C_{\mathrm{sat}}$ at the interface,  the initial bubble-front distance ($H_0 \approx \SI{125}{\micro \meter}$, see App.\,\ref{appB}),  and the advancing velocity of the front (see inset figure\,\ref{fig:4}($c$)).
We obtain $\Delta C$ by fitting the slopes of the experimental data according to (\ref{Eq:EpsteinPlesset}). 
Rescaling by the size of the bubble as it touches the ice, $R_{\mathrm{bub,\infty}}$, and the typical diffusion time, $t_{\mathrm{diff}} = \rho_g R_{\mathrm{bub},\infty}^2 / (2 D \Delta C)$,  shows that the initial bubble growth for all cases indeed follows (\ref{Eq:EpsteinPlesset}) (see figure\,\ref{fig:4}($c$)). 
However, as the bubble grows and the front approaches, its growth deviates from that of pure diffusion under initial conditions and is enhanced.

To pinpoint the origin of the enhanced bubble growth, we perform numerical simulation to study their dynamics under various conditions (see \S\,\ref{sec:numsim} for technical details).
We ensure that all physical parameters are known from the literature or obtained experimentally, and that we do no rely on any adjustable fitting parameters. 
Initially, advection is omitted and we only consider growth through diffusion. The effect of advection around the bubble is addressed in more detail in App.\,\ref{appC}.

For the numerical simulations we consider three distinct cases.
Firstly, we take the case of a bubble growing in a homogeneous background concentration, with the front, located at a distance $H_0 = \SI{125}{\micro \meter}$, remaining static.
We retrieve a behaviour very similar to (\ref{Eq:EpsteinPlesset}) (see black line figure\,\ref{fig:6}), where the deviation at the end is a consequence of the presence of the static interface, bounding the domain, and slightly altering the concentration field around the bubble (see panel I in figure\,\ref{fig:6}).  
Secondly, we assume a background concentration gradient according to (\ref{Eq:ConcentrationProfile}), resulting in a significantly faster growth of the bubble (see blue line and panel II in figure\,\ref{fig:6}).
Thirdly, we consider the front to be in motion with velocity $V_{\mathrm{front}}$ and to approach the growing bubble, which further enhances its growth (see green line and panel III in figure\,\ref{fig:6}).
The determined bubble growth of the latter agrees very nicely with our experimental observations,  where the faster growth originates from a combination of the background gas concentration gradient and the motion of the interface, causing an accumulation of the iso-concentration lines at the bottom of the bubble (see panel III in figure\,\ref{fig:6}).
This alters the local concentration gradient close to the interface of the bubble, leading to an enhanced mass transfer and thus a more rapid growth. 

\subsection{Enhanced mass transfer}
\label{sec:mass transfer}

\begin{figure}
  \centering
  \includegraphics[width=0.45\textwidth]{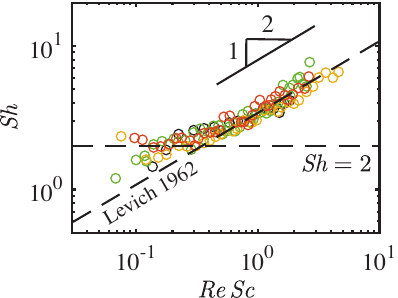}
  \caption{$\Sh$ versus $\Reyn \, \Sc$ for different bubbles growing near the advancing solidification front, represented by different colors. 
The data tends towards the limits $\Sh = 2$ and $\Sh \sim \left (Re \, \Sc \right )^{1/2}$ (dashed black lines) \citep{levich1962physicochemical}.}
\label{fig:5}
\end{figure}

To further rationalise the observed enhanced mass transfer, we recast our problem of a bubble growing near an advancing solidification front into that of the well-studied case of mass (or equivalently heat) transfer to a moving spherical object in an infinite domain with a constant background concentration (or temperature).
The dimensionless diffusional flux to the moving object can then be expressed as \citep{levich1962physicochemical}
\begin{equation}
\Sh \propto \left( \Reyn \, \Sc \right)^{1/2}
\label{Eq:Levichflux}
\end{equation}
where the dimensionless Sherwood, Reynolds, and Schmidt numbers are defined as
\begin{equation}
\Sh = \frac{2R_{\mathrm{bub}} \dot{R}_{\mathrm{bub}} \rho_g}{D \Delta C}, \quad \Reyn = \frac{\rho U R}{\mu}, \quad \text{and} \quad \Sc = \frac{\mu}{\rho D},
\label{Eq:DimensionlessNumbers}
\end{equation}
with $\dot{R}_{\mathrm{bub}} = \mathrm{d}R_{\mathrm{bub}}/\mathrm{d}t$ the growth rate of the bubble and $U$ the relative fluid velocity at the equator of the spherical object.
Whereas the Sherwood number is defined as the ratio of the convective mass transfer rate to diffusive mass transport, the product of Reynolds and Schmidt is equivalent to a Peclet number, \textit{i.e.}, $\Reyn \, \Sc = \Pe$, which is defined as the ratio of the rate of advection to that of diffusion.
In the limit that $\left(\Reyn \, \Sc \right) \rightarrow 0$ the solution to this classical case is $\Sh = 2$, where the concentration field look similar to that depicted in panel I of figure\,\ref{fig:6}. 
As the object moves through the (homogeneous) concentration field, the iso-concentration lines tend to accumulate at one end \citep{steinberger1960mass,frossling1963evaporation,ihme1972theoretische, oellrich1973theoretische}, similar to panel III of figure\,\ref{fig:6}, leading to an enhanced mass transfer according to (\ref{Eq:Levichflux}).

In our case, the bubble is stationary and the background concentration is not constant. 
To account for the moving background concentration gradient the bubble experiences, we introduce an effective velocity as
\begin{equation}
U_{\mathrm{eff}} = \left\vert \frac{\mathrm{d} C}{\mathrm{d} y} \right\vert \frac{R }{C_0} V_{\mathrm{front}}.
\label{Eq:Ueff}
\end{equation}
Considering that $U = U_{\mathrm{eff}}$, we obtain that our experimental observations are in line with (\ref{Eq:Levichflux}) as we approach the limit $\left(\Reyn \, \Sc \right) \rightarrow \infty$ (see dashed black line in figure\,\ref{fig:5}).
In the opposite limit, when the bubble and the background concentration gradient are both small, \textit{i.e.}, $\left(\Reyn \, \Sc \right) \rightarrow 0$,  the system undergoes a transition towards a different scaling, approaching $\Sh \approx 2$, where the notable scatter arises from inaccuracies in determining the bubble growth rate $\dot{R}_{\mathrm{bub}}$ for smaller bubbles. 

\section{Conclusions \& outlook}
\label{sec:conclusion}

To conclude, we have investigated the freezing of a quasi-two-dimensional sessile drop on a cooled substrate. 
During the freezing process, gases dissolved in water are rejected by the growing ice crystal and accumulate at the moving ice front, creating a supersaturated region which favours the nucleation of gas bubbles.
These bubbles then grow, before eventually being engulfed into the ice.
Their final shape and size in the ice is governed by a delicate interplay between the rate of freezing and the rate of mass transfer towards the bubble \citep{bari1974nucleation,lipp1987investigation}. 

In this work, we focussed on the experimental and numerical investigation of the initial growth of those bubbles that nucleate not at but near the solidification front.
We show that at the early stages the growth is dominated by diffusion and is enhanced due to a combination of the presence of the background gas concentration gradient and the motion of the approaching front. 
We rationalised that, based on the numerical simulations, which agree nicely with the experimental observations, the iso-concentration lines tend to accumulate at the bottom of the bubble, leading to an enhanced mass transfer across the interface of the bubble, and hence a faster growth. 
We have additionally shown, that our problem can be recast into that of mass transfer to a moving spherical object in a homogeneous concentration field, finding good agreement between our experimental data and the existing scaling relations.
Finally,  through numerical simulations we have qualitatively addressed how fluid flow around the bubble might further affect its growth (see App.\,\ref{appC}).

Our findings shed new light on the diverse processes that might govern the growth of bubbles near a moving interface, subjected to gradients in both concentration \textit{and} temperature.
Besides any solidification process of pure or multi-component liquids, such as metals, silica or sapphires, where these findings might be relevant, they might also bring new insight into the formation of gas bubbles in hailstones \citep{bari1974nucleation} and lake ice \citep{swinzow1966ice,gow1977growth} under various conditions. 
As continuation, it might be interesting to investigate how the bubble growth is altered when freezing liquids that are saturated with a more (less) soluble gas, such as CO$_2$ (Ar), to further explore the relevance of our findings.

\section*{Acknowledgements}
The authors thank Gert-Wim Bruggert, Martin Bos and Thomas Zijlstra for the technical support and Maaike Sikkink and Iris Kormelink for preliminary experiments.
The authors acknowledge the funding by Max Planck Center Twente, the Balzan Foundation, and the support by an Industrial Partnership Programme of the Netherlands Organisation for Scientific Research (NWO) \& High Tech Systems and Materials (HTSM),  co-financed by Canon Production Printing Netherlands B.V., University of Twente, and Eindhoven University of Technology.
This work was part of JM’s Ph.D. dissertation \citep{meijer2024particles}.

\section*{Declaration of Interest}
The authors report no conflict of interest. 

\appendix
\section{Bottom and drop temperatures during freezing}\label{appA}

\begin{figure}
  \centering
  \includegraphics[width=\textwidth]{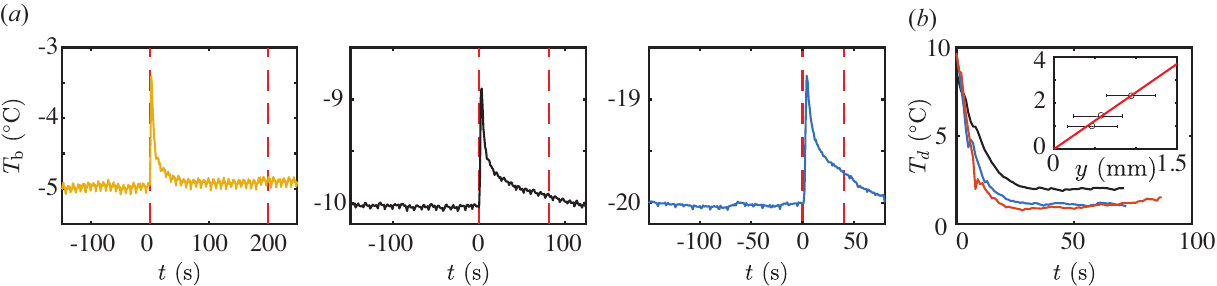}
  \caption{($a$) Measured bottom temperature $T_b(t)$ underneath the deposited (at $t=0$) drop, which initially had room temperature, for three different substrate temperatures. During the freezing process (indicated by the red dashed lines) $T_b(t)$ deviates from the fixed substrate temperature $T_s$. ($b$) Temperature $T_d$ within the droplet measured at three fixed distance when $T_s = \SI{-10}{\celsius}$.  }
\label{fig:SI_1}
\end{figure}

\begin{figure}
  \centering
  \includegraphics[width=0.85\textwidth]{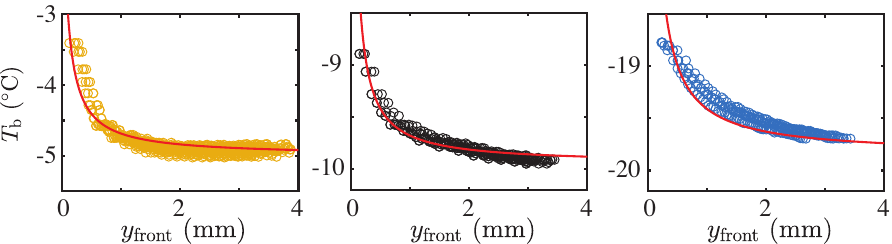}
  \caption{Measured bottom temperature $T_b(t)$ underneath the deposited drop as a function of the position of the ice front, $y_{\mathrm{front}}(t)$ (see figure\,\ref{fig:3}), for three different substrate temperatures.  The solid red lines correspond to (\ref{Eq:Tb}) with fitted values for $\lambda$.  We have $\lambda = \SI{2.4e4}{\watt \per \meter \squared \per \kelvin},\SI{7.9e4}{\watt \per \meter \squared \per \kelvin},\SI{9.7e4}{\watt \per \meter \squared \per \kelvin}$ for $T_s = \SI{-5}{\celsius}, \SI{-10}{\celsius}, \SI{-20}{\celsius}$, respectively. }
\label{fig:SI_3}
\end{figure}

While measuring the bottom temperature $T_b$ during the freezing process, we observe a rapid increase in temperature when the drop, which is of room temperature, is deposited.
After this deposition, the measured temperature slowly decays back towards the set substrate temperature (see figure\,\ref{fig:SI_1}($a$)). Using the second equation of (\ref{Eq:FrontEquationsFull}) we can derive an expression of $T_b$ as a function of $y_{\mathrm{front}}$ as
\begin{equation}
T_b  = \frac{k_i T_m + \lambda T_s y_{\mathrm{front}}}{k_i + \lambda y_{\mathrm{front}}},
\label{Eq:Tb}
\end{equation}
with $k_i = \SI{2.2}{\watt \per \meter \per \kelvin}$ the thermal conductivity of ice, $T_m$ its melting temperature, $T_s$ the set substrate temperature and $\lambda$ a (fitted) heat-transfer coefficient. 
Combining the results of figure\,\ref{fig:3}\,($a$) with figure\,\ref{fig:SI_1}($a$), and fitting (\ref{Eq:Tb}) through the experimental data (see figure\,\ref{fig:SI_3}), we find $\lambda = \SI{2.4e4}{\watt \per \meter \squared \per \kelvin},\SI{7.9e4}{\watt \per \meter \squared \per \kelvin},\SI{9.7e4}{\watt \per \meter \squared \per \kelvin}$ for $T_s = \SI{-5}{\celsius}, \SI{-10}{\celsius}, \SI{-20}{\celsius}$, respectively.

Additionally, to have a rough estimate on the temperature within the drop as it solidifies, we use a thermocouple immersed in the drop to measure the liquid temperature at three fixed distances from the advancing front (see figure\,\ref{fig:SI_1}($b$)) when $T_s = \SI{-10}{\celsius}$.  
This allows for a first-order estimation of the temperature gradient. We obtain $\mathrm{d}T_d/\mathrm{d}y \approx \SI{2.5e3}{\kelvin \per \meter}$ (see inset figure\,\ref{fig:SI_1}($b$)).

\section{Concentration and surface tension profiles at the front}\label{appB}

\begin{figure}
  \centering
  \includegraphics[width=\textwidth]{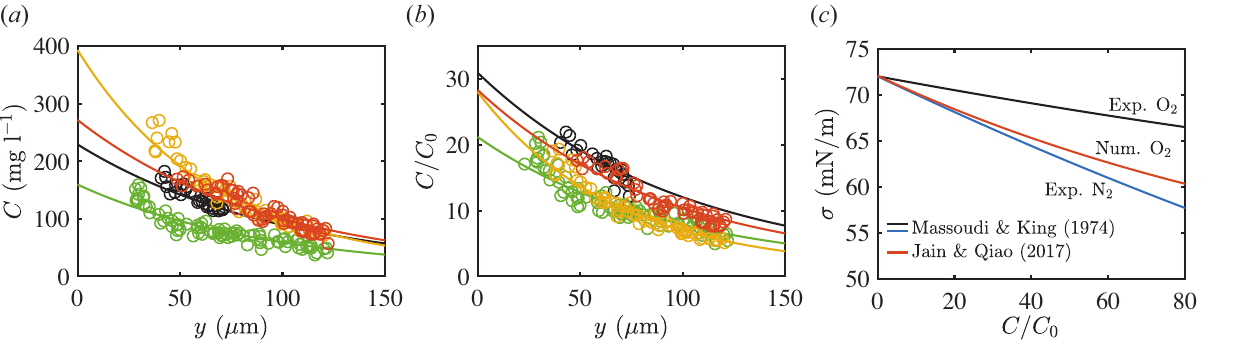}
  \caption{($a$) Gas concentration $C$ as a function of distance y to the advancing front, for different bubbles, represented by different colors, as determined by (\ref{Eq:deltaC_exp}). ($b$) Gas supersaturation $C/C_0$ in water as a function of distance. ($c$) Surface tension $\sigma$ as function of the supersaturation for O$_2$ and N$_2$ in water \citep{massoudi1974effect,jain2017molecular}. }
\label{fig:SI_2}
\end{figure}

In order to determine the concentration profile at the solidification front, we rewrite (\ref{Eq:EpsteinPlesset}) to obtain \citep{bari1974nucleation,lipp1987investigation} 
\begin{equation}
\Delta C = C(y) - C_{\mathrm{sat}} =  \frac{\rho_g R_{\mathrm{bub}}}{D}\frac{\mathrm{d}R_{\mathrm{bub}}}{\mathrm{d}t}, \quad \text{with} \quad C_{\mathrm{sat}} = \lambda_{\mathrm{N}_2}\left(p_{\mathrm{atm}} + \frac{2 \sigma}{R_{\mathrm{bub}}}\right),
\label{Eq:deltaC_exp}
\end{equation}
where $\lambda_{\mathrm{N}_2} = \SI{1.7e-7}{\kg \per \meter \cubed \per \pascal}$ is Henry's law constant for N$_2$ in water.  
We relate the time-dependence $R(t)$ to a position-dependence $R(y)$ by tracking the center of mass altitude $H(t)$ of the bubble with respect to the approaching front (see figure\,\ref{fig:4}($d$)).
From this we obtain our experimental result for $C(y)$, see figure\,\ref{fig:SI_2}($a$).
The initial, diffusive growth is fitted with (\ref{Eq:ConcentrationProfile}) in order to determine the partition coefficient ($K = 0.037 \pm 0.007$) and the far field gas concentration in water ($C_0 = \left(9 \pm 3\right) \, \SI{}{\mg \per \litre}$).
This allows us to then determine the gas supersaturation $C/C_0$ at the front (see figure\,\ref{fig:SI_2}($b$)).
The dependency of surface tension on the gas supersaturation is taken from the literature and shown in figure\,\ref{fig:SI_2}($c$).

\section{Effect of fluid flow around the bubble}\label{appC}

{\renewcommand{\arraystretch}{1.3}
\begin{table}
\centering
\begin{tabular}{llll}
Parameter & Value & Parameter& Value \\
$\langle R_{\mathrm{bub}} \rangle \, (\SI{}{\meter})$ & $25 \times 10^{-6}$ (see figure\,\ref{fig:4}(\textit{b})) &$\Ra_T$&$\approx 3 \times 10^{-4}$ \\
$\mathrm{d}\rho/\mathrm{d}T  \, (\SI{}{\kg \per \meter \cubed \per \kelvin})$ & $\approx 0.05$ for $\SI{0}{\celsius} < T < \SI{4}{\celsius}$ &$\Ra_C$& $\approx 2 \times 10^{-3}$\\
$\mathrm{d}\rho/\mathrm{d}C $ &  \begin{tabular}[c]{@{}l@{}} O$_2$: $3.1 \times 10^{-2}$ [1] \\ N$_2$: $-0.232$ [1,2] \end{tabular}  &$\Ma_T$&$\approx 56$\\
$\mathrm{d}\sigma/\mathrm{d}T \, (\SI{}{\newton \per \meter \per \kelvin})$ &  $-1.3 \times 10^{-4}$&$\Ma_C$&$\approx 6$ \\
$\mathrm{d}\sigma/\mathrm{d}C \, (\SI{}{\meter \cubed \per \second \squared})$ & \begin{tabular}[c]{@{}l@{}} Num. O$_2$: $-1.95 \times 10^{-2}$ ([3], see App.~\ref{appB}) \\ Exp. O$_2$: $-7.7 \times 10^{-3}$ ([4], see App.~\ref{appB})\\ Exp. N$_2$: $-1.99 \times 10^{-2}$ ([4], see App.~\ref{appB})\end{tabular}  && \\
$\mathrm{d}T/\mathrm{d}y  \, (\SI{}{\kelvin \per \meter})$ & $\approx 2.5 \times 10^{3}$ (see App.~\ref{appA}) && \\
$\mathrm{d}C/\mathrm{d}y \, (\SI{}{\kg \per \meter \tothe{4}})$ & $\approx -3 \times 10^{3}$ (see App.~\ref{appB}) && \\
$\mathrm{d}C_{\mathrm{sat}}/\mathrm{d}T \, (\SI{}{\kg \per \meter \tothe{3} \per \kelvin})$ & $\approx -7 \times 10^{-4}$ && \\
\end{tabular}
\caption{Physical parameters and their values, either taken from the literature or determined experimentally. They are necessary to compute the dimensionless numbers introduced in equations \,(\ref{Eq:Rayleigh})\&(\ref{Eq:Maragoni}).
[1]: \citep{watanabe1985influence} [2]: \citep{soto2019transition} [3]: \citep{jain2017molecular} [4]: \citep{massoudi1974effect}}
\label{tab:my-table}
\end{table}
}


In the main part of the paper we have considered that the bubble growth is governed by diffusion, which tends to account well for our experimental observations.
Nonetheless, given the configuration of our system, in principle flow around a bubble at an advancing solidification front might emerge in the form of solutal/thermal natural convection \citep{enriquez2014quasi,dietrich2016role, soto2019transition} or solutal/thermal self-induced Marangoni advection \citep{young1959motion, li2019bouncing, li2021marangoni,li2022marangoni, zeng2021periodic,meijer2023rising}.
Due to the presence of both concentration and temperature gradients and the dependency of mass density and surface tension on each, four potential origins of such flow arise: thermal or solutal natural convection, or thermal or solutal Marangoni flow.  
In this appendix we explore whether any of such flow could be of relevance for the growing bubble at the ice front. We will conclude this appendix that this is not the case.

Thermal and solutal natural convection would give rise to an \textit{upwards} flow, since the colder/more N$_2$-saturated and hence lighter liquid is located at the moving front, \textit{i.e.}, $\left[\left(\mathrm{d}\rho / \mathrm{d}T \right)  \left(\mathrm{d} T / \mathrm{d}y\right), \left(\mathrm{d}\rho / \mathrm{d}C_{N_2} \right)  \left(\mathrm{d} C/ \mathrm{d}y\right) \right]  > 0$ (see table\,\ref{tab:my-table}).
To quantify their importance with respect to the diffusive process we define a thermal and solutal Rayleigh number as the ratio of the natural convection time-scale to the diffusion time-scale \citep{li2022marangoni}
\begin{equation}
\Ra_T = \frac{g R^4}{\mu D} \frac{\mathrm{d} \rho}{\mathrm{d}T}\frac{\mathrm{d} T}{\mathrm{d}y},  \quad \text{and} \quad \Ra_C = \frac{g R^4}{\mu D} \frac{\mathrm{d} \rho}{\mathrm{d}C}\frac{\mathrm{d} C}{\mathrm{d}y}, 
\label{Eq:Rayleigh}
\end{equation}
where $g$ is gravitational acceleration and $\mu$ the dynamic viscosity of water. Other physical parameters and their corresponding values are tabulated in table\,\ref{tab:my-table}.  
When evaluating these Rayleigh numbers for our specific case, it becomes apparent that diffusion dominates over natural convection and $\Ra_T$ and $\Ra_C$ are very small, see table\,\ref{tab:my-table}. 

We now turn to the self-induced Marangoni advection.  
Whereas the thermal Marangoni advection around a bubble is well understood \citep{young1959motion} and would cause a \textit{downwards} flow, \textit{i.e.}, $\left(\mathrm{d}\sigma / \mathrm{d}T \right)  \left(\mathrm{d} T/ \mathrm{d}y\right)  < 0$ (see table\,\ref{tab:my-table}), its solutal counter-part requires a more elaborate discussion. 
Considering thermodynamic equilibrium at the bubble interface, the local concentration of the dissolved gases is set by the saturation concentration, $C_{\mathrm{sat}}$, governed by Henry's law \citep{yang2018marangoni, massing2019thermocapillary}.  
Given the temperature dependence of the Henry coefficient, the thermal gradient present in the water (see App.\,\ref{appA}) might induce an opposing \textit{upwards} solutal Marangoni flow \citep{lubetkin2003thermal}, since the saturation concentration decreases with increasing temperature, \textit{i.e.}, $\left(\mathrm{d} C_{\mathrm{sat}} / \mathrm{d}T \right)  < 0$, and surface tension decreases with increasing concentration, \textit{i.e.}, $\left(\mathrm{d} \sigma / \mathrm{d}C \right) < 0$, leading to $\left(\mathrm{d} T / \mathrm{d}y \right) \left(\mathrm{d} C_{\mathrm{sat}} / \mathrm{d}T \right) \left(\mathrm{d} \sigma / \mathrm{d}C \right) > 0$ (see table\,\ref{tab:my-table}).
Additionally, surface active species such as dissolved gases \citep{lubetkin2002motion}, surfactants \citep{meulenbroek2021competing} or ions \citep{park2023solutal} might also induce concentration gradients along the interface, altering the surface tension locally and hence inducing a solutal Maragoni flow.  
We can compare the ratio of the advection time-scale, due to the self-induced Maragoni flows, to the diffusive time-scale, expressed as two Maragoni numbers \citep{li2022marangoni}
\begin{equation}
\Ma_T = \frac{V_{M,T} R}{D} =  -\frac{1}{2}\frac{R^2}{\mu D} \frac{\mathrm{d} \sigma}{\mathrm{d}T}\frac{\mathrm{d} T}{\mathrm{d}y}, \quad \text{and} \quad \Ma_C = \frac{V_{M,C} R}{D} = \frac{1}{2}\frac{R^2}{\mu D} \frac{\mathrm{d} \sigma}{\mathrm{d}C}\frac{\mathrm{d} C_{\mathrm{sat}}}{\mathrm{d}T} \frac{\mathrm{d} T}{\mathrm{d}y},
\label{Eq:Maragoni}
\end{equation}
where $V_M$ is the self-induced thermal ($T$) or solutal ($C$) Marangoni velocity at the equator of the bubble \citep{young1959motion,li2022marangoni}.
The physical parameters and their corresponding values are tabulated in table\,\ref{tab:my-table}.
Although the determined values of the Marangoni numbers are moderate (also see table\,\ref{tab:my-table}), efforts to experimentally visualise the flow around the growing bubbles through particle tracking velocimetry (PIV) did not yield valuable insights at the time.     

\begin{figure}
  \centering
  \includegraphics[width=0.95\textwidth]{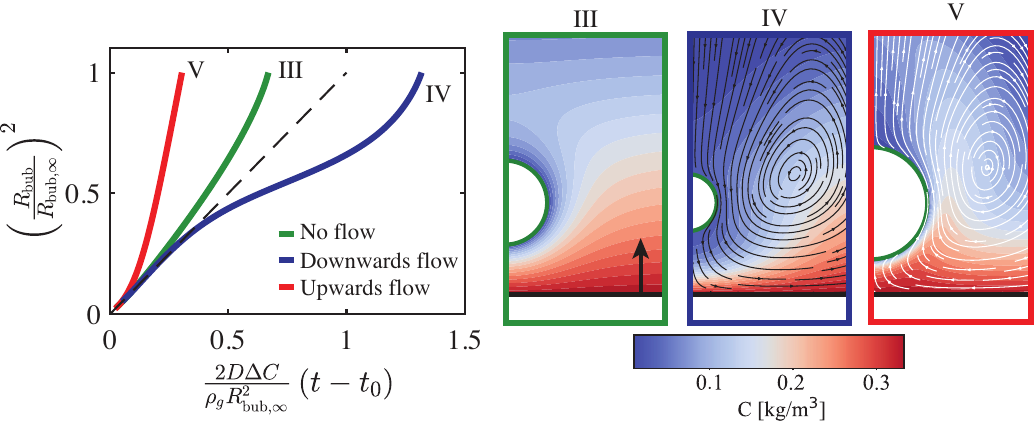}
  \caption{Numerical simulations (Suppl. Movie 6) on the growth of a bubble near an advancing solidification front considering downwards (IV) and upwards (V) advection at the interface of the bubble.  The panels show the streamlines of the flow superimposed on the concentration profiles around the bubbles, 3 seconds after their initialisation. White-colored (black-colored) streamlines indicate (anti-)clockwise rotation. The initial bubble radius is $R_0 = \SI{10}{\micro \meter}$, $C_{\mathrm{sat}} = \SI{0.029}{\kg \per \meter \cubed}$, and $V_{\mathrm{front}} = \SI{10}{\micro \meter \per \second}$ (for all cases).}
\label{fig:7}
\end{figure}

Alternatively, we once again turn to the numerical simulations to obtain qualitative insights on how fluid flow around the growing bubble might affect its growth.
By altering the strength of the Marangoni advection artificially, we are able to generate flow at the interface of the growing bubble in upwards and downwards direction. 
These imposed flow structures alter the local concentration profile in the vicinity of the bubble and therefore hinder or accelerate its growth, compared to the purely diffusive case discussed earlier (panel III in figure\,\ref{fig:6}). 
Whereas the emergence of a downwards flow retards the bubble growth, as advection depletes the supersaturated region around the bubble (see panel IV in figure\,\ref{fig:7}), the growth is accelerated when the flow is upwards. 
The liquid in the vicinity of the bubble becomes more and more enriched, leading to a significantly more rapid growth (see panel V in figure\,\ref{fig:7}).
This is in contrast to our experimental observations, and therefore we think that also Marangoni flows do not play a role in the bubble growth process under investigation.

\bibliographystyle{jfm}
\bibliography{References}

\end{document}